# First demonstration of a bubble-assisted Liquid Hole Multiplier operation in liquid argon

E. Erdal*, A. Tesi, A. Breskin, D. Vartsky and S. Bressler

*Department of Particle Physics and Astrophysics, Weizmann Institute of Science, Rehovot 7610001, Israel*

*Email*: eran.erdal@weizmann.ac.il

**Abstract**

ABSTRACT: We demonstrate, for the first time, the operation of a bubble-assisted Liquid Hole Multiplier (LHM) in liquid argon. The LHM, sensitive to both radiation-induced ionization electrons and primary scintillation photons, consists of a perforated electrode immersed in the noble liquid, with a stable gas-bubble trapped underneath. Electrons deposited in the liquid or scintillation-induced photoelectrons emitted from a photocathode on the electrode's surface, are collected into the holes; after crossing the liquid-gas interface, they induce electroluminescence within the bubble.

After having validated in previous works the LHM concept in liquid xenon, we provide here first preliminary results on its operation in liquid argon. We demonstrate the bubble containment under a Thick Gas Electron Multiplier (THGEM) electrode and provide the detector response to alpha particles, recorded with SiPMs and with a PMT - under electroluminescence and with modest gas multiplication; the imaging capability is also demonstrated.

* Corresponding author

**Contents**

**1. Introduction**

The concept of Liquid Hole-Multipliers (LHMs) was proposed [1] for the combined detection of ionization electrons and primary scintillation photons generated along charged-particle tracks in noble liquids [2-7]. The original goal was to deploy LHM units within a single-phase noble liquid TPC, aiming at simplifying the structure and thus overcoming potential technological issues that may occur in large-volume dual-phase TPC detectors. The latter are the leading instruments in current dark-matter searches [8] and are part of the future neutrino-physics program [9, 10].

A conceptual scheme of a bubble-assisted LHM module is depicted in Figure 1. It consists of a perforated electrode (e.g., a gas electron multiplier (GEM) [11], or a thick gas electron multiplier (THGEM) [12]) immersed in the noble liquid, with a stable gas-bubble trapped underneath; the latter is formed by an array of thin heating wires. Once formed, the bubble fills the space below the electrode and remains stable as long as the system is in a thermodynamic steady state. Radiation-induced electrons deposited in the liquid or scintillation-induced photoelectrons emitted from a CsI photocathode deposited on the electrode's top surface, are collected into the electrode's holes; after crossing the liquid-gas interface, they generate electroluminescence (EL) within the bubble. The LHM electrodes are held at different potentials, assuring the efficient collection of ionization electrons and scintillation-induced photoelectrons into the electrode's holes and their extraction into the bubble - where they induce EL; the resulting photons are detected with photo-sensors located underneath.

In our LHM-based "*local dual-phase*" TPC concept (Figure 1), a prompt scintillation signal (S1) is followed by two EL ones. The first, S1', results from S1 photons, extracting photoelectrons from CsI into the bubble; the second one, S2, results from ionization electrons focused into the holes. While the time difference between S1 and S1' is constant, S2 arrives at a time depending on the "depth" of interaction (dictated by the electrons' drift velocity in the liquid).

The new concept has been validated in series of works carried out in liquid xenon (LXe) [2-7]. Best results in terms of light yield and energy resolution were reached with a single-conical GEM perforated electrode: EL yield ~400 photons/$e^-$/4 and energy resolution ~5% RMS (for ~7000 alpha-induced ionization electrons in LXe) [6]. Imaging of an alpha-particle pattern performed with a quad-SiPM coupled to a THGEM-LHM, permitted its reconstruction with a resolution of ~200 µm RMS [7].

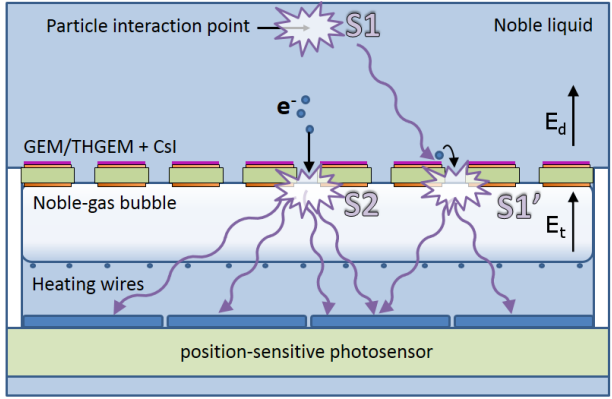

*Figure 1 Conceptual scheme of a LHM detector. A hole-electrode (GEM/THGEM) with CsI photocathode deposited on its top surface is immersed in the noble liquid. Heating wires generate a vapor bubble underneath the electrode; the bubble is laterally supported by two walls. Ionization electron drift towards the electrode and generate EL signals. Scintillation photons extract a photoelectron from the CsI layer and induce similar EL signal. The EL signals are read using position sensitive light readout pads (PMTs, SiPM etc.).*

In this work, we report for the first time, on the operation of a bubble-assisted LHM in liquid argon (LAr), at a temperature of ~90 K. We discuss the operation of a THGEM-based LHM and present preliminary examples of alpha-induced signals, energy spectra and imaging capabilities.

It is worth mentioning that EL in THGEM holes immersed in LAr has already been reported in [13]; however, in light of the present knowledge, we suspect this might have been due to sporadic bubbles accumulating under the electrode, rather than to EL in the liquid as reported by the authors. Similarly, electron multiplication has been reported on arrays of sharp tips immersed in LAr [14]. In another work [15, 16], pressure-dependent components of avalanche multiplication in LAr with small Xe admixtures were suspected to arise from Ar bubbles forming at the edge of the sharp tips.

## 2. Experimental setup & methodology

The experiments were conducted in a dedicated LAr cryostat, WISArD (Weizmann Institute Liquid Argon Detector). It comprises a 100 mm in diameter, 150 mm tall cylindrical chamber filled with ~250 ml of LAr. The rest of the volume is equipped with various sensors, with the detector assembly suspended from the topmost flange; the latter has a viewport.

Argon liquefaction is done on cooling fins cooled by a Cryomech PT90 pulse tube refrigerator. Temperature of the liquefaction fins is maintained by a closed-loop feedback using CryoCon temperature controller (Model 24). The liquefied argon flows in through a 1.5 m long double-wall, vacuum-insulated transfer tube into the chamber. During operation, LAr is continuously extracted through a tube immersed within the liquid, evaporated and circulated through an SAES hot getter (PS3-MT3-R) at 2-5 standard liters per minute. The purified Ar gas returns to the liquefaction fins where it is liquefied and is let to flow back into the chamber.

The LHM detector assembly (see Figure 2) is almost identical to the one described in [7]. It consists of a 0.4 mm thick THGEM electrode (hexagonal holes pattern 0.3mm in diameter, 0.7 mm apart). In this proof-of-principle study, the THGEM electrode was not coated with CsI, making the detector sensitive only to ionization electrons. Note that although a THGEM is not the best electrode in terms of light yield and of photon detection efficiency (as described in details in [6]), it was chosen for this first demonstration for its mechanical and electrical robustness. A $^{241}$Am alpha source was placed 4.7 mm above the THGEM electrode. The source had an activity pattern of an annulus ~0.5 mm broad and ~3.9 mm in diameter with total activity of ~190 Bq (see [7] for details). A grid of heating wires (Ni-Fe, 55 µm in diameter, 2 mm spacing) was placed 1.6 mm below the THGEM. It was used to generate the bubble and to define the "transfer field" ($E_t$), i.e. the field between the bottom face of the THGEM and the wires. For convenience, we will quote the values of the transfer field as if it was a parallel plate between the bottom face of the electrode and the heating wires; this is only indicative, as the actual electric field is non uniform, being high close to the bottom part of the THGEM hole and next to the wires [6] and lower in the gap between them.

Since the current cryostat does not have a side window allowing to visually observe the bubble (as in previous LXe work, e.g. [4]), the existence of the bubble was inferred from the response of the EL signals to sudden pressure changes (as detailed in [3]).

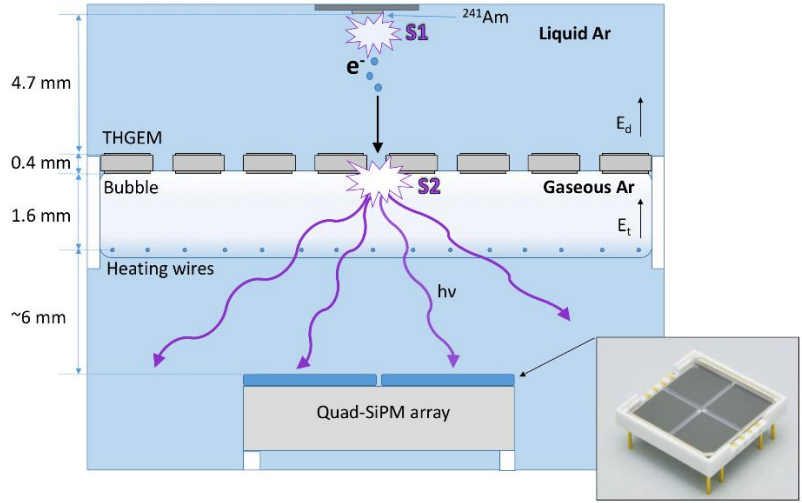

*Figure 2 Schematic view of the experimental setup. A 0.4 mm thick THGEM electrode (with 0.3 mm in diameter holes spaced by 0.7 mm) is immersed in LAr; the gas bubble underneath is formed by a grid of heating wires, spaced 2 mm apart, located 1.5 mm below the electrode. Photons were recorded either by a PMT or by a Quad-SiPM array (shown here), located at ~6 mm under the wire grid. Ionization electrons focused into the holes induce EL light (S2) in the bubble; a fraction of the primary scintillation photons (S1) traverses the holes and is detected as well.*

Readout of the EL photons was done by direct digitization of the signals from the photosensors, located ~6 mm below the wire grid, using a Tektronix (MSO 5204B) oscilloscope. Two photosensors were used: (1) a PMT (Hamamatsu R8520-506) vacuum-coated with ~300 µg/cm² of TPB (Tetraphenyl butadiene) wavelength shifter and operated at -800 V. The PMT has a fast response and therefore allows for signal shape reconstruction. (2) A four-elements windowless SiPM (Hamamatsu VUV4 MPPC S13371-6050CQ-02, total area 12x12 mm²); the quad-SiPM permitted 2D event-position reconstruction. The SiPM was operated at -46V as per the manufacturer's recommendation. However a full study of its operation at cryogenic temperatures has not yet been performed. Therefore, the resolutions quoted here are preliminary and may improve in future work.

The recorded waveforms were analyzed using dedicated Matlab scripts. Position of the event was reconstructed by a center-of-gravity technique as described in [7].

## 3. Results

Prior to biasing of the THGEM electrode, current was driven through the heating wires in order to generate a bubble (30V across 61Ω for ~10 seconds). Once formed, no further heating was applied to the wires. As previously shown [3], the EL signals disappeared upon sudden rise in the pressure and reappeared immediately after its sharp decrease. While not being able to visually see the bubble in the present setup, this confirms its existence. All measurements where conducted at a liquid temperature of 90K, corresponding to a pressure of 1365 mbar.

### 3.1 Typical signals

With the LHM detector polarized to $\Delta V_{THGEM} = 3,000 V$ (keeping $E_d = 1\ kV/cm$ and $E_t = 0\ kV/cm$), typical signals as recorded with the TPB-coated PMT are shown in Figure 3; Figure 4 depicts signals recorded from the four SiPMs' pads. The green marked pulses are that of the primary scintillation light (S1); the red ones originate from the EL photons from the bubble (S2). One can observe a longer S2 decay constant (3-4 µs) compared to that in LXe (e.g. see [6]). There are indeed different processes in LAr, with longer time scales than in LXe. First, Ar has two decay constants for the scintillation or EL signal, 5 ns and 860 ns, as opposed to Xe where the longest timescale is 27 ns [17]. Then, the transition of electrons from liquid to gas has, according to [18], both long and short timescales. None of them, however, can explain the 3-4 µs decay constant, observed with the two different sensors. This matter will be part of further studies.

### 3.2 Energy resolution

For each event, the pulse area under each waveform was computed and a pulse-area histogram was plotted. An example recorded with $\Delta V_{THGEM} = 3,000\ V\ E_d = 1\ kV/cm$ and $E_t = 0\ kV/cm$, recorded with the PMT at liquid temperature of 90° K is shown in Figure 5. A Gaussian fit was applied for deriving the resolution and the mean value.

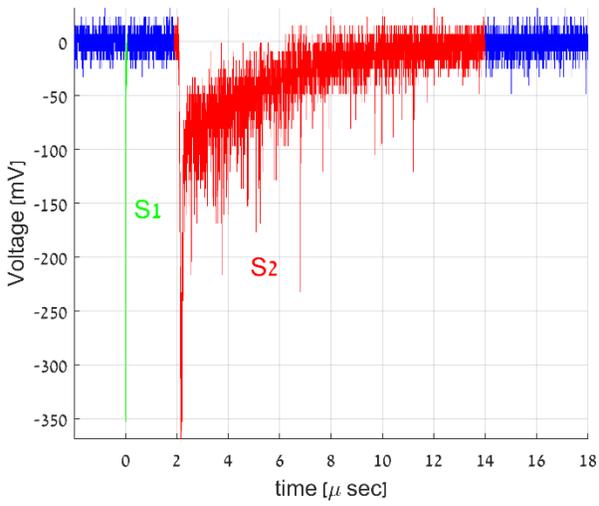

*Figure 3 Example of alpha-particle induced single-event waveform, recorded by the TPB-coated PMT.*

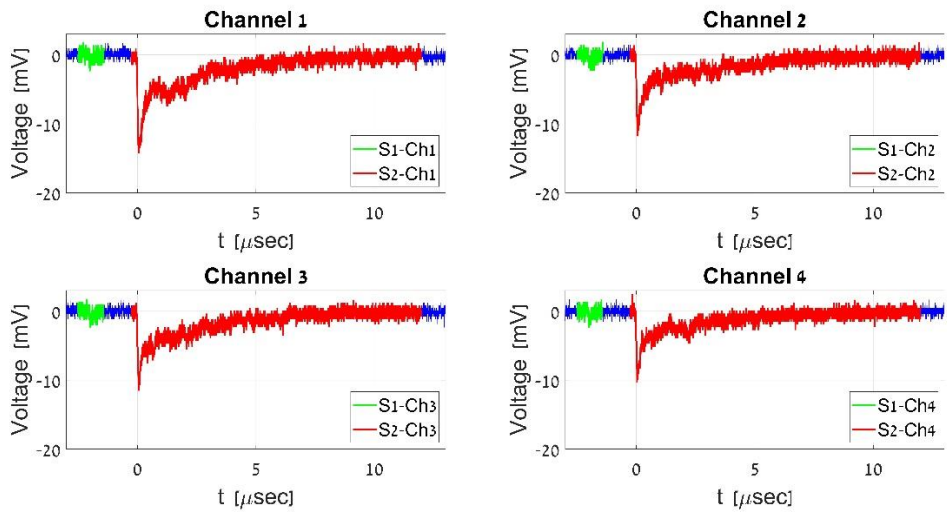

*Figure 4 Example of alpha-particle induced single-event waveforms, recorded in the LAr-LHM setup of Fig. 2, by the four SiPM pads. $V_{SiPM}$=-46V*

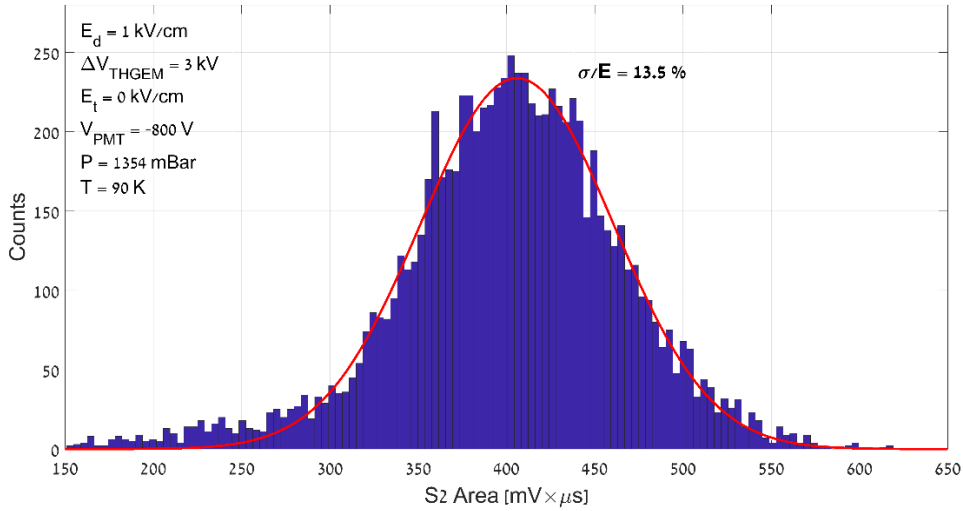

*Figure 5 Area distribution of alpha-particle induced EL pulses recorded with the PMT, in the LAr-LHM setup of Fig. 2, at $\Delta V_{THGEM} = 3,000\ V\ E_d = 1\ kV/cm$ and $E_t = 0\ kV/cm$, at temperature of 90°K. A Gaussian fit was applied to the data, for deriving the mean value and the RMS resolution.*

Figure 6a shows the pulse-area as a function of the voltage across the THGEM electrode (keeping $E_d = 1\ kV/cm$ and $E_t = 0\ kV/cm$) and Figure 6b shows the RMS resolution. The linear trend shown in Figure 6a indicates EL without charge gain [17]. The RMS resolution was measured to be 13.5%. This value is ~2-fold worse than the one achieved in LXe and requires additional studies. Similar response to that measured with the TPB-coated PMT was confirmed with the windowless quad-SiPM.

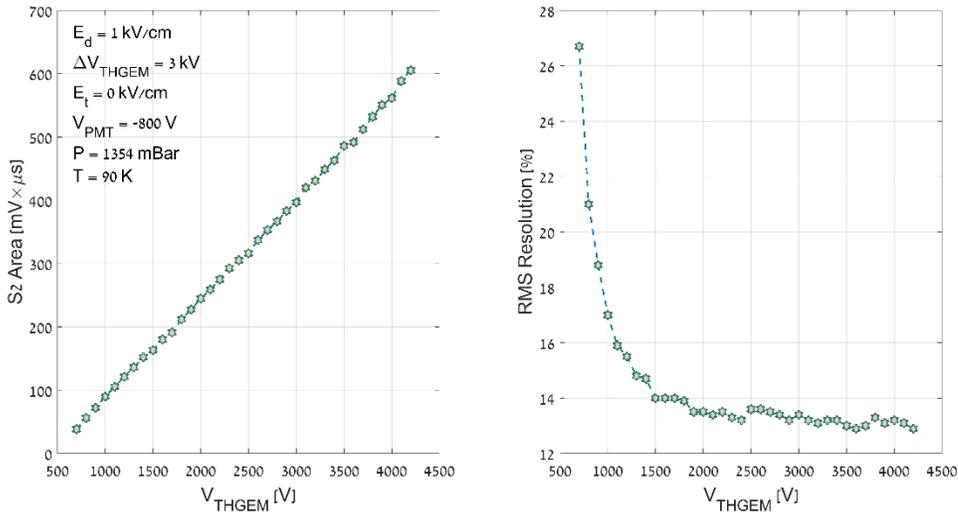

*Figure 6 Response of the PMT to alpha particles, of the LAr-LHM detector of Fig. 2. a) S2 mean area as a function of the voltage across the THGEM electrode. (b) RMS resolution of the area distribution.*

## 3.3 Amplification in the transfer gap

As discussed in details in [3] and in [6], the light yield can be boosted by increasing the electric field between the bottom of the THGEM electrode and the heating wires beneath ($E_t$). At moderate values of $E_t$ (~2 kV/cm), this results in EL generated at the vicinity of the heating wires in addition to the one occurring mainly at the bottom of the holes. At more intense fields, the electrons generate EL all along their path within the bubble: from the bottom of the hole to the wires. Figure 7 shows the average pulse-shape obtained at different $E_t$ values. One can clearly see the addition of a second EL pulse, occurring ~0.5 µsec after that originating from the vicinity of the THGEM hole; it results in a significant increase in signal magnitude.

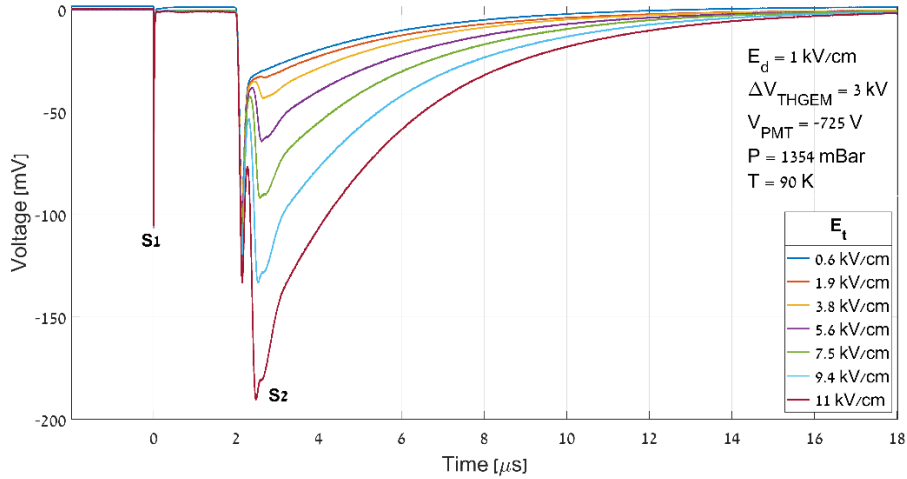

*Figure 7 Average pulse shape recorded by the PMT from the LAr-LHM detector, at different values of the transfer field, $E_t$. $\Delta V_{THGEM} = 3,000\ V$ and $E_d = 1\ kV/cm$.*

An example of the resulting pulse-area spectrum recorded with the PMT with $E_t = 15$ kV/cm is shown in Figure 8. It is interesting to observe that at such high $E_t$ value, pulses resulting from the 59.5 keV gamma particles (emitted by the $^{241}$Am) become visible. The data was recorded here with $V_{PMT} = -725\ V$ to avoid signal saturation. One should notice that although the gamma- and the alpha-particles differ in their energy by almost two orders of magnitude, their response in the LAr-LHM detector differ only by a factor ~4.5. This is due to the known different recombination probabilities of the ionization electrons, between the dense alpha-induced ionization and sparse density of electrons induced by the gamma-rays [17, 19].

The peak position the Gaussian fit is depicted in Figure 9, as a function of $E_t$ . Once can clearly see that above ~4 kV/cm, the pulse-area grows exponentially, indicating upon modest (~10-fold) charge-avalanche multiplication at the vicinity of the wires.

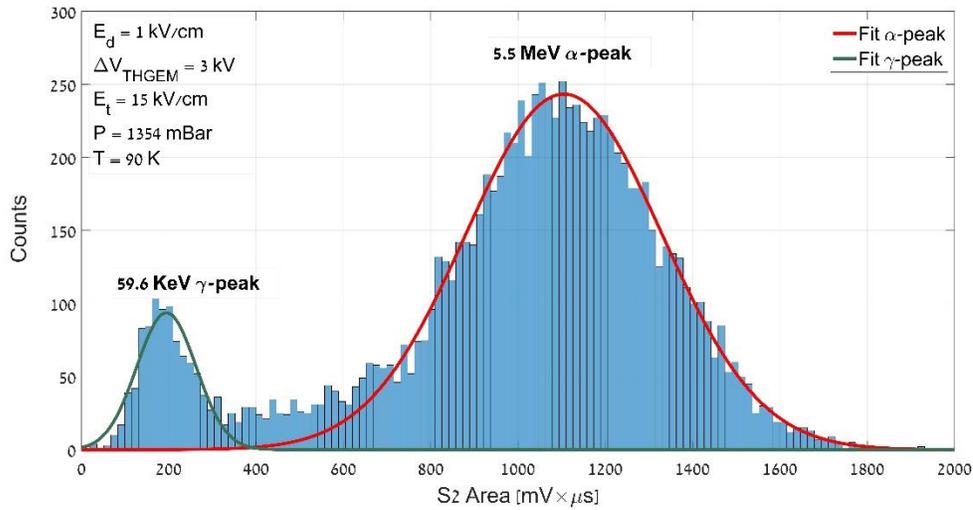

*Figure 8 Pulse-area spectrum recorded by the PMT from the LAr-LHM detector with EL occurring in the THGEM holes, in the transfer gap and near the heating wires. One Gaussian fit (red) corresponds to the 5.5 MeV alpha particles and the second one (green), to the 59.5 keV gamma interactions in LAr.*

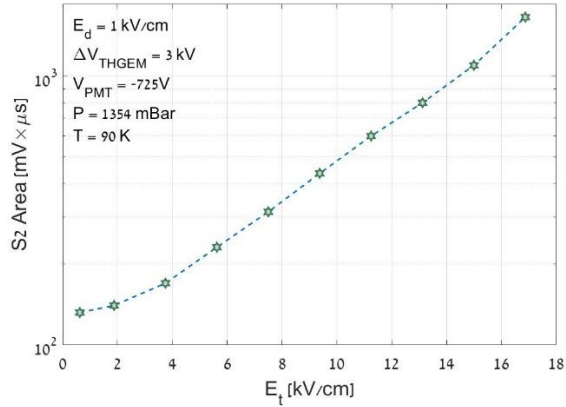

*Figure 9 Signal magnitude of the alpha particle peak of Figure 7, as a function of the transfer field.*

### 3.3 Position reconstruction

Similar to the methodology presented in [7], the integral of the pulse from each of the four SiPM pads was computed for each event. The event position was then reconstructed by a center-of-gravity method. An auto-radiographic image (using a Fuji phosphor-imager scanner model FLA-9000, plate model BAS-TR2040S) of the $^{241}$Am alpha source used in our experiments is shown in Figure 10a. The 2D histogram of the derived event positions, recorded in LAr with the LHM (of Figure 2), is shown in Figure 10b. This very preliminary qualitative image reproduces the annular shape of the alpha source. However, the reconstruction resolution is, at this point, poorer in comparison to the ~200 µm RMS one, recorded with LXe-LHM [7], calling for further investigation.

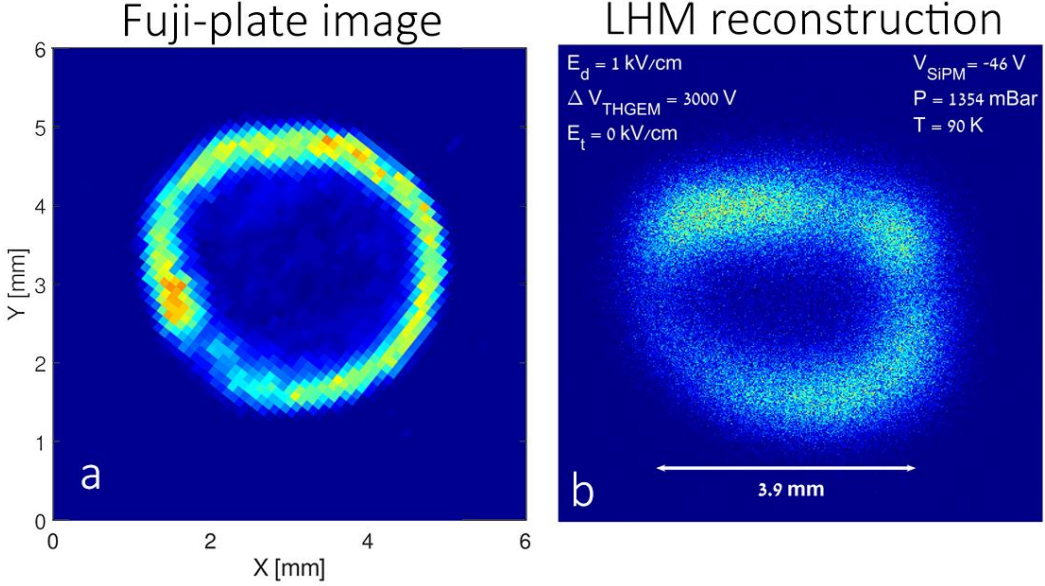

*Figure 10 (a) The auto-radiographic Alpha-source image, recorded with a Fuji phosphor-imaging plate. (b) 2D histogram of the EL photons emitted at the vicinity of the liquid-gas interface, recorded with the Quad-SiPM LAr-LHM detector.*

## 4. Summary and discussion

In this work, we have demonstrated, for the first time, the operation of an LHM detector in LAr. We have shown that similar to LXe, a bubble can be generated in LAr and is sustained for a long period under a THGEM perforated electrode immersed in the liquid. Electroluminescence (EL) within the Ar bubble, induced by ionization electrons deposited by alpha particles in LAr and collected into the electrode's holes, shows a linear response with the applied voltage, as expected in such process. Electrons drifting within the bubble towards the heating-wires grid induced modest charge-avalanche multiplication. Imaging of the alpha-particle induced EL photons was demonstrated, qualitatively, with the quad-SiPM.

This preliminary demonstration paves ways towards a more elaborate and quantitative study of the bubble-assisted LHM detector in Ar. It will encompass the optimization of the perforated electrode geometry, study of the physical processes governing electron transfer through the liquid-gas interface, optimization of the readout (SiPM, PMT etc.) and of other parameters. We thus expect enhancing the (yet non-calibrated) EL light yield and improving the localization resolution over a broader ionization range. The combined detection of ionization signals and scintillation photons, with CsI-coated perforated electrodes, will be investigated in detail.

Although at its early stages of research and development, it is planned to investigate the merits of the LHM concept as a potential tool in future large-volume noble-liquid experiments. E.g., the LHM-based TPC is part of the R&D program of DARWIN – the future LXe-based dark matter experiments [20]. The LAr-based LHM could also be considered in the context of many other fields where LAr is the technology of choice, including Dark Matter searches [21], neutrino physics [9] as well as neutron detection in the contexts of future nuclear-physics experiments (e.g. [22]), homeland security [23] and many more.

## 5. Acknowledgements

We would like to thank Dr. M. L. Rappaport (Weizmann Institute of Science - WIS) and to Dr. H. Wang (UCLA) for their invaluable advices with the LAr cryostat-system design and Dr. A. Roy (WIS and Ben Gurion University) for his assistance with the LAr cryostat setup. We also thank Dr. N. Canci (INFN-LNGS), assisted by M. Weiss (WIS) for the TPB deposition. We are indebted to O. Diner, H. Sade, A. Jahanfard, Y. Asher and Y. Chalaf (all from WIS) for their technical design and production. This work was partly supported by the Israel Science Foundation (Grant No. 791/15) and by the I-CORE Program of the Planning and Budgeting Committee. The research was carried out within CERN-RD51 detector R&D research program.